\documentclass[12pt,preprint]{aastex}
\usepackage{graphicx}
\newcommand{\met}{\hbox{E\kern-0.5em\lower-0.1ex\hbox{/}}_T}
\newcommand\simlt{\lower.5ex\hbox{$\; \buildrel < \over \sim \;$}}
\newcommand\simgt{\lower.5ex\hbox{$\; \buildrel > \over \sim \;$}}

\begin{document}
\title{Recollimation and Radiative Focusing of Relativistic Jets: Applications to Blazars and M87}
\author{Omer Bromberg \& Amir Levinson\altaffilmark{1}}
\altaffiltext{1}{Raymond and Beverly Sackler School of Physics \& Astronomy, Tel Aviv University,
Tel Aviv 69978, Israel; Levinson@wise.tau.ac.il}


\begin{abstract}
Recent observations of M87 and some blazars reveal violent activity in small regions located 
at relatively large distances from the central engine.  Motivated by these considerations, 
we study the hydrodynamic collimation of a relativistic cooling outflow
using a semi-analytical model developed earlier.  We first demonstrate that radiative 
cooling of the shocked outflow layer can lead to a focusing of the outflow and its 
reconfinement in a region having a very small cross-sectional 
radius.  Such a configuration can produce rapid variability at large distances from the central  engine via 
reflections of the converging recollimation shock.  Possible applications of this model to TeV blazars 
are discussed.  We then apply our model to M87.  The low radiative efficiency
of the M87 jet renders focusing unlikely.  However, the shallow profile of the ambient medium pressure inferred
from observations, results in extremely good collimation that can explain the reported variability of the X-ray 
flux emitted from the HST-1 knot.
\end{abstract}
\keywords{radiation mechanism: non-thermal - shock waves - galaxies:jets - galaxies:active - galaxies:individual: M87}

\section{Introduction}
The interaction of a relativistic jet with the surrounding matter is likely to play an important role in its collimation
and the dissipation of its bulk energy.  Such an interaction is evident on the largest scales, where  
hot spots, presumably associated with termination shocks, and features that appear to be 
associated with recollimation by the surrounding matter and by backflows from the jet endpoints
are often observed in AGNs and, in some cases, microquasars.  Less clear is the effect of the environment on 
relativistic jets at much smaller scales, in particular in the blazar zone. 

Recent observations of TeV AGNs suggest that a considerable fraction of the bulk energy may dissipate in 
reconfinement shocks on VLBI scales. These observations motivate reconsideration of the standard view, 
according to which the broad-band, highly variable emission seen in blazars is produced predominantly behind 
internal shocks that form in colliding fluid shells.  One indication is the systematic differences  between
the large values of the Doppler factor inferred from TeV observations (e.g., Levinson 2006; Begelman et al. 2008)  
and the much lower values inferred from unification models (Urry and Padovani 1991; Hardcastle et al. 2003) and 
superluminal motions on parsec scales (e.g., Marscher 1999; Jorstad et al. 2001).  Such differences can be naturally 
accounted for if dissipation occurs in quasi stationary patterns, e.g., oblique shocks, that result
from the interaction of the jet with ambient matter.  In this scenario radio observations reflect the pattern speed
while the Lorentz factor inferred from TeV observations is associated with the speed of the fluid passing the structure.
Furthermore, the large Doppler factors implied by opacity arguments may not be required in the first place
if the TeV emission is produced at radii larger than commonly believed.  
However, in that case channeling of the bulk energy into a rather small area needs to be assumed in order 
to account for the rapid variability often observed.  Alternative explanations have also been offered in order to resolve the
so called 'Doppler factor crises', including a structure consisting of interacting spine and sheath (Ghisellini et al. 2005), opening 
angle effects (Gopal-Krishna 2004) and jet deceleration (Georganopoulos \& Kazanaz 2003; Levinson 2007).

Based on analysis of multi-wavelength observations during the 2005 outburst in 3C454.3, Sikora et al. (2008) argued that the 
blazar emission in this source is produced at a distance of several parsecs from the putative black hole, where the mm 
photosphere is located.  They proposed that the blazar activity is driven by a standing 
reconfinement shock that, they claim, dissipates energy more efficiently than internal shocks.  They further argued
that the high energy emission is likely produced via inverse Compton scattering of IR dust emission, and demonstrated
that the observed SED can be reconstructed in such a model.   A fit of their model 
to the observed broadband spectrum favors a relatively large Lorentz factor of the emitting fluid, $\Gamma\sim20$, consistent 
with that required to account for the temporal variations of the optical and millimeter fluxes.  

Perhaps the best example that an outflow energy can be channeled into a small area located far from the central engine is 
the HST-1 knot in M87, a stationary radio feature associated with the sub-kpc scale jet.  The knot 
is located at a projected distance of 60 pc ($0.86''$) from the central engine, and is known to be 
a region of violent activity.   Sub-features moving away from the main knot of the HST-1 complex 
at superluminal speeds have been detected recently (Cheung et al. 2007).  In addition, rapid, large amplitude variations of 
the resolved X-ray emission from HST-1 have been reported, with a doubling time $t_{\rm var}$ of $0.14$ yrs.
The observed variability limits the linear size of the X-ray source to
$d \simlt \delta_D ct_{\rm var}\sim 0.044\delta_D$ pc, where $\delta_D$ is the Doppler factor, 
which for a viewing angle of $\theta_n\sim30^{\circ}$
is at least three orders of magnitude smaller than the distance between the HST-1 knot and the central black hole.
The constraints on the size of the radiation source are much more severe if the observed TeV emission also 
originated from HST-1, as suggested by some based on a claimed correlation between the X-ray and TeV emission 
(e.g., Cheung et al. 2007, but c.f., Neronov \& Aharonian 2007). 
It has been proposed that HST-1 is associated with a recollimation nozzle (Stawarz et al. 2006; hereafter ST06).  
In this picture the rapid variability and superluminal sub-knots that seem to be expelled from the HST-1 complex can 
be associated with shocks produced by reflection of the recollimation shock at the axis (Levinson and Bromberg 2008).  The 
rapid variability sets a limit on the cross-sectional radius of the channel at the location of HST-1 that depends on the 
fraction of jet power radiated as X-rays (and TeV emission, if indeed originating from the same location).  The conditions required
to produced a structure consistent with this picture are examined in \S 3 below.

In what follows we exploit a model developed by us earlier to address some of the issues discussed above.  
We generalize it to include radiative losses and demonstrate that even modest radiative cooling of the 
shocked jet material can lead to extremely good focusing.  The effect of cooling on the collimation and 
confinement of non-relativistic jets has been studied earlier in the context of SS433 (e.g., Peter \& Eichler 1995). 
Our work presents an extension of this idea into the relativistic regime.
A preliminary account of the results presented below is outlined in Levinson and Bromberg (2008).

\section{The Model}
In a previous paper (Bromberg and Levinson 2007; hereafter BL07) we constructed a class of semi-analytical models 
for the confinement and collimation of a relativistic jet by the pressure and inertia of a surrounding medium.
Both, confinement by kinetic pressure of a static corona, and confinement by the ram pressure of a supersonic 
wind emanating from a disk surrounding the inner source have been considered.  
In general, the collision of the inner jet with the confining medium leads to
the formation of a contact discontinuity across which the total pressure
is continuous, and an oblique shock across which the streamlines of the colliding flow are deflected.
In cases where confinement of the inner flow is accomplished through collision with a supersonic wind
a second shock forms in the exterior wind.  
The model outlined in BL07 computes the structure of 
the shocked layers of the deflected inner jet and the exterior wind in the latter case, assuming a 
steady, axisymmetric flow.  In BL07 the focus was on the application to GRBs.  Radiative losses have 
been ignored since the large optical depth of the shocked jet layer on scales of interest renders such losses negligibly small.
In blazars, recollimation shocks are expected to form above the photosphere.  If the 
cooling rate of the shocked plasma is high enough, then a significant fraction of the thermal energy may 
be radiated away and this may affect the structure of the 
flow.  To study this effect we incorporated radiative losses into our model.  The results are described below. 

\subsection{Analytical approximations}
Consider the interaction of a relativistic outflow with a gaseous condensation extending from some fiducial height 
$z_0$ above the equatorial plan to infinity, and having a pressure profile $p_{ext}(z)=p_0 (z/z_0)^{-\eta}$.  The 
outflow is assumed to be ejected from a point source into a cone of opening 
angle $\theta_{j}$, with a total power $L_j$ distributed uniformly inside the cone, and velocity $\beta_{j0}$ at $z=z_0$.
As explained above, the structure of the confined flow consists of a contact discontinuity separating the 
shocked jet layer and the ambient medium and a collimation shock \footnote{For certain external pressure profiles the flow
equations admit self-similar solutions with no shocks (Zakamska et al., 2008).}.
The details of this structure will depend, quite generally, on the parameters of
the injected outflow and on the external pressure.   We denote by $r_c(z)$ and $r_s(z)$ 
the cross-sectional radii of the contact discontinuity and shock surfaces, respectively.  By employing the 
shock jump conditions and requiring  momentum balance at the contact discontinuity surface, the  hydrodynamic 
equations can be reduced to a coupled, nonlinear set of ODEs for $r_c(z)$, $r_s(z)$ and the thermodynamic parameters
(density, temperature, etc.) of the shocked fluid (BL07).  In general, solutions must be sought numerically.
However, the problem can be considerably simplified in certain regimes.  At small angles one obtains to 
lowest order (Komissarov \& Falle 1997; BL07),
\begin{equation}
\frac{dr_s}{dz}=\frac{r_s}{z}-A\tan\theta_jz^{1-\eta/2},
\label{shc-evl-diff}
\end{equation}
the solution of which is 
\begin{equation}
r_s(z)=z\tan\theta_j-\frac{2A}{2-\eta}z\tan\theta_j\left(z^{1-\eta/2}-z_0^{1-\eta/2}\right),
\label{shc-evl}
\end{equation}
where $A=(\pi c p_0z_0^\eta/L_j\beta_{j0}\xi_1)^{1/2}$. Here $\xi_1=(1-{\hat n}\cdot{\bf\beta}_{+}/{\hat n}\cdot{\bf\beta}_-)\simlt1$, with ${\bf\beta}_{-}$ and ${\bf\beta}_{+}$ denoting the local fluid 3-velocity upstream and downstream of the shock, respectively, and ${\hat n}$ the normal to the shock surface.  BL07 adopted $\xi_1=1$ for convenience whereas ST06 adopted $\xi_1=0.7$.  The point $z^\star$ at which the shock reaches the axis is determined from the condition $r_s(z=z^\star)=0$.
Using eq. (\ref{shc-evl}) one finds,
\begin{equation}
z^\star=\left(\frac{2-\eta}{2A}+z_0^{1-\eta/2}\right)^{1/(1-\eta/2)}.
\label{zstar}
\end{equation}
Note that $z^\star$ depends solely on the total jet power $L_j$ and the external pressure profile. 
However, as will be shown below, the jet profile depends also on the opening angle and other details.  
From eq. (\ref{zstar}) it is readily seen that convergence is guaranteed for $\eta\le2$.  In the case $\eta>2$  
the shock will approach the axis provided $p_0$ exceeds some critical value $p_c$, given by
\begin{equation}
p_c=\frac{(\eta-2)^2L_j\beta_{j0}}{4\pi c z_0^2}.
\label{pc}
\end{equation}
As $p_0$ approaches $p_c$ the location of the reflection point $z^\star$ approaches infinity, and when $p_0<p_c$ the shock diverges.  
The structure of the polar outflow in the latter case consists of a core containing the unshocked jet enveloped by the shocked jet 
layer that expands relativistically, but with a Lorentz factor considerably smaller than that of the unshocked jet,
as described in detail in BL07.  

Approximate solutions can also be obtained analytically for the contact discontinuity surface.  In the absence of radiative energy 
losses the radius $r_c(z)$ at $z>>z_0$ is given approximately by (BL07)
\begin{equation}
r_c(z)=r_{c0}(z/z_0)^{\eta/4}.
\label{cont-evl}
\end{equation}
Evidently, collimation occurs when $\eta<4$, however, the flow never shrinks, even in 
situations where the shock itself reaches the axis.   The reason is that the pressure inside the shocked layer 
pushes against the ambient gas and keeps it expanding as long as $\eta>0$.
The situation may change if the shocked gas cools radiatively at a high enough rate that allows strong compression
shocked jet material.  In that case, the shocked layer remains thin, leading to the convergence of the contact 
discontinuity towards the axis.  In the limit of extremely
rapid cooling the width of the shocked layer approaches zero, viz., $r_c\simeq r_s$, and focusing is expected under the same
conditions that lead to convergence of the collimation shock.   
\begin{figure}[h]
\centering
\includegraphics[width=11cm]{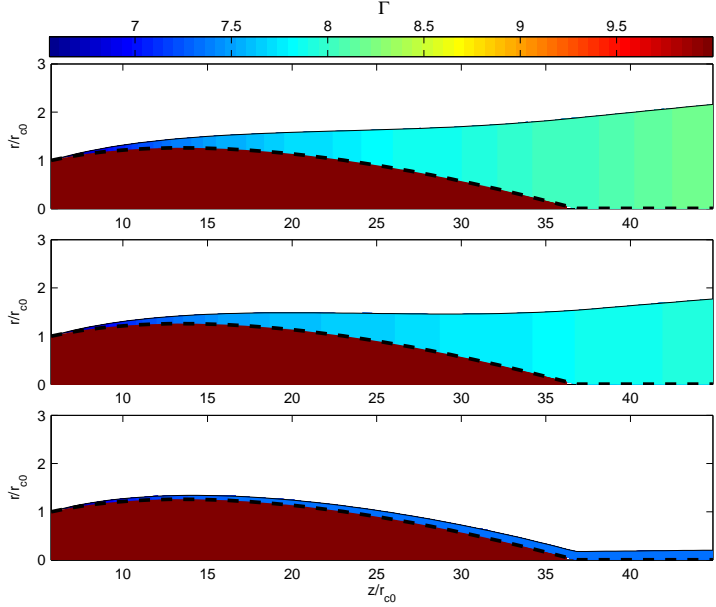}
\caption{\small{Lorentz factor of the shocked and unshocked jet in the case of confinement by a 
static corona with a pressure profile $p_{ext}\propto z^{-2}$, for different values of the cooling parameters.
In all cases shown $\xi_B=0.01$.
The ratio of the total luminosity radiated away by the shocked jet layer and total jet power 
is $L_c/L_j=0$ (no radiative losses) in the upper panel,
$L_c/L_j=0.1$ ($\xi_e=0.015$) in the middle panel, and $L_c/L_j=0.27$ ($\xi_e=0.06$) in the lower panel.
The injected flow in all panels consists of a cold, purely baryonic fluid with
Lorentz factor $\Gamma_{j0}=10$ at the injection point $z=z_0$.  The shock surface is marked by the dashed line}}
\end{figure}

\begin{figure}[h]
\centering
\includegraphics[width=11cm]{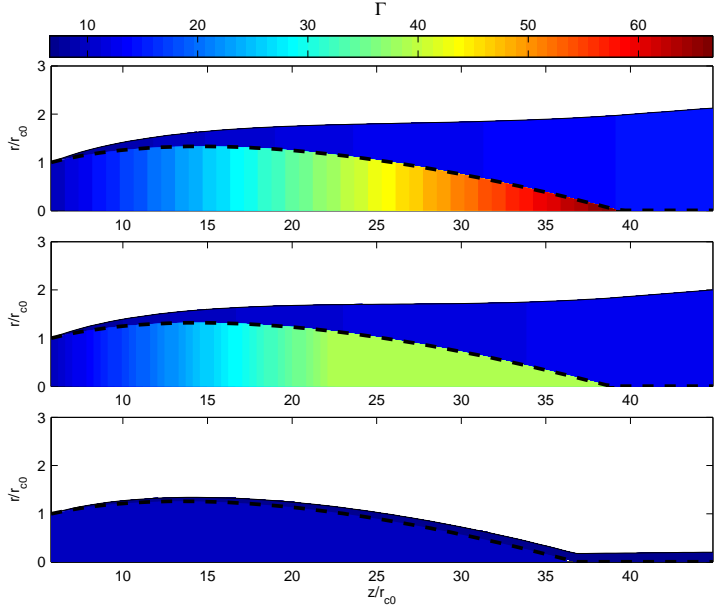}
\caption{\small{The dependence of the solution on jet temperature.  The external pressure profile 
and the value of $\xi_B$ are the same as in fig. 1.  The parameter $\xi_e$ was adjusted such that 
$L_c/L_j=0.27$ in all panels.   The power of the injected flow is given to a good approximation by 
$L_j=h^\prime\dot{M}_j\Gamma c^2$, where $h^\prime$ denotes
the proper entropy per baryon of the unshocked fluid.
The ratio of total jet energy to rest mass energy is $L_j/\dot{M}_j\Gamma c^2=h^\prime=10$ in the upper panel, 
$L_j/\dot{M}_j\Gamma c^2=4$ in the middle panel, and $L_j/\dot{M}_j\Gamma c^2=1$ in the lower panel.}}
\end{figure}

\begin{figure}[h]
\centering
\includegraphics[width=11cm]{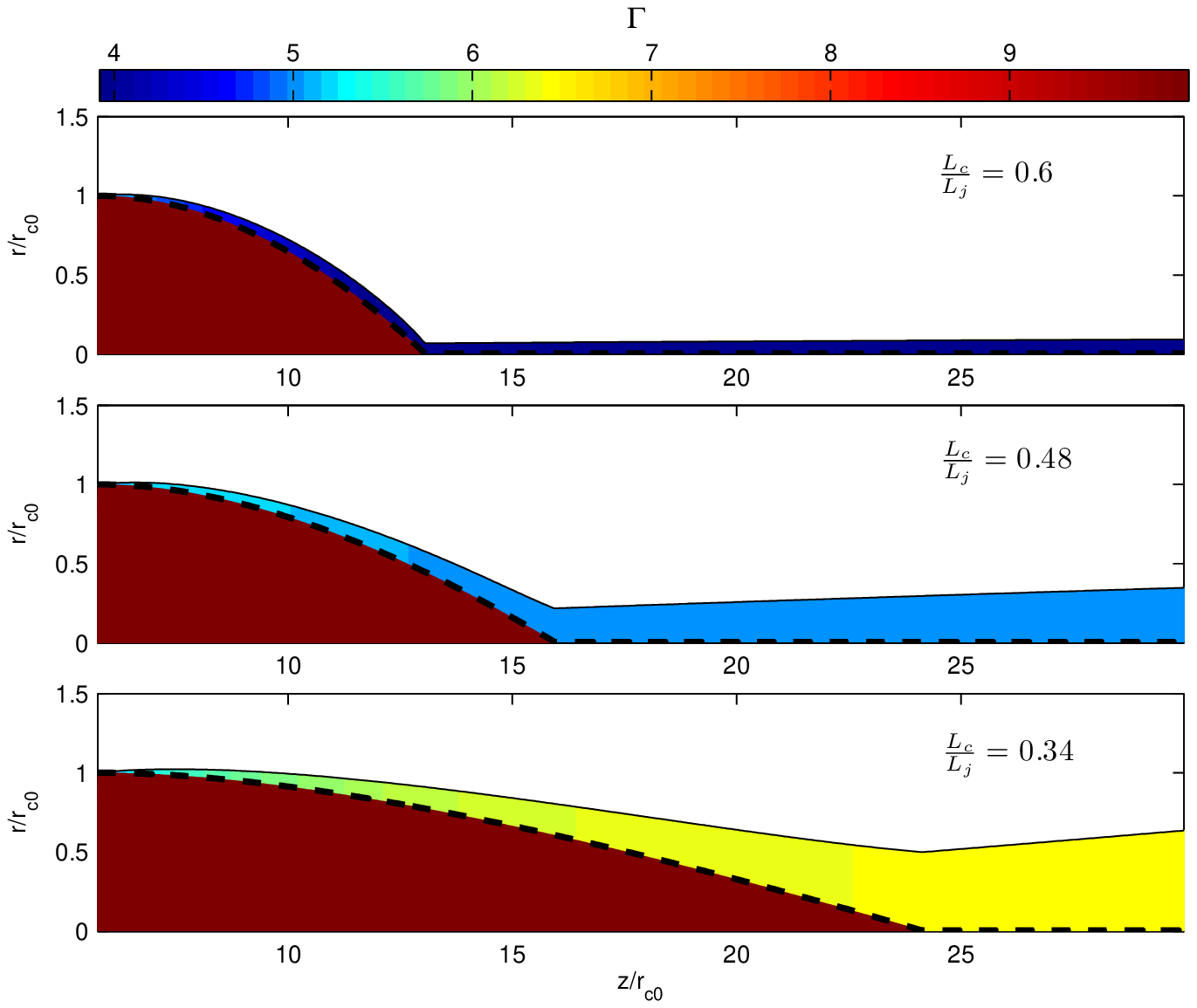}
\caption{\small{The effect of the ambient medium.  In all cases shown $\Gamma_{j0}=10$, $L_j/\Gamma\dot{M}_jc^2=1$, $\xi_B=0.01$,
and $\xi_e=0.04$.  The upper, middle and lower panels correspond to $\eta=1$, 2, and 3, respectively.
The radiative efficiency $L_c/L_j$ is indicated in each panel.}}
\end{figure}

\subsection{Radiative reconfinement and  focusing of a relativistic flow}
A fraction of the shocked energy is likely to be tapped for acceleration of particles to nonthermal energies, e.g., by 
shock acceleration, or shear mechanism (Ostrowski, 1998).  The thermal and nonthermal 
electrons may cool via synchrotron and/or inverse Compton emission, leading to compression of the shocked layer.   
This compression is clearly independent of the details of the cooling mechanism, but merely on the fraction of the total  
kinetic energy (the sum of thermal energy and the energy carried by the nonthermal population) that is being lost to 
radiation over dynamical scales.  
We suppose that a fraction $\xi_B$ of the total energy flux in the shocked layer is carried by magnetic fields.
The parameter $\xi_B$ can be readily related to the sigma parameter of the injected outflow via the shock jump conditions.
We consider sufficiently low sigma flows for which the magnetic field is dynamically unimportant.  Let us assume first that
the shocked plasma is in a thermal state, and that electrons and protons are strongly coupled (that is, the equilibration 
time is much shorter than the flow time).  If the specific enthalpy in the shocked layer is dominated by rest mass,
as we find to be the case for cold, baryonic jets, then to a good approximation $U^\prime_B=B^{\prime 2}/8\pi= \xi_B m_pc^2n^\prime_p$
(the prime refers to quantities measured in the rest frame of the flow).
Since the shocked layer is confined by the external pressure the average energy of thermal electrons is $\epsilon^\prime_e=
\gamma^\prime_em_ec^2=p_{ext}/2n^\prime_e$.   With $n^\prime_e=n^\prime_p$ and $p_{ext}=p_0(z/z_0)^{-\eta}$, the ratio of cooling time due to synchrotron and inverse Compton emission,
$t^\prime_{cool}=5\times10^8B^{\prime-2}\gamma^{\prime-1}_e(1+U^\prime_{ph}/U_{B}^\prime)^{-1}$ s, where $U^\prime_{ph}$ is the 
energy density of the background radiation field measured in the fluid rest frame, and flow time, $t^\prime_f=z/c\Gamma$, 
is given by:
\begin{equation}
\frac{t^\prime_{cool}}{t^\prime_{f}}=2\left(1+\frac{U^\prime_{ph}}{U_{B}^\prime}\right)^{-1}\left(\frac{\xi_B}{0.1}\right)^{-1}
\left(\frac{P_0}{10^{-2} {\rm \ dyn \  cm^{-2}}}\right)^{-1}\left(\frac{z_0}{0.1 \ {\rm pc}}\right)^{-1}
\left(\frac{z}{z_0}\right)^{\eta-1}\Gamma.
\label{th-cool}
\end{equation}
Thus, cooling of the thermal plasma on timescale shorter than the flow time requires 
$U^\prime_{ph}/U_{B}^\prime$ $\simgt(\xi_B/0.1)^{-1}\Gamma$ for
the above choice of parameters.  If a fraction $\tau_d$ of the disk luminosity $L_d$ is scattered across the jet by the surrounding
gas (e.g., Blandford and Levinson 1995), then $U_{ph}\simeq \tau_dL_d/4\pi c r^2$, and $U^\prime_{ph}=\Gamma^2U_{ph}$ on account of
beaming of the radiation field in the comoving frame.  Rapid cooling of thermal electrons then implies 
$L_d/L_j\simgt0.1\tau_d^{-1}\Gamma^{-3}\theta_j^{-2}$, where $\theta_j$ is the opening angle of the jet, and may be expected 
in ERC blazars.

It could well be that a small fraction of the electrons in the shocked layer are being continuously picked up 
from the thermal pool and injected to nonthermal energies.  This population cools rapidly 
and can significantly enhance the radiative loss of the shocked layer.
For instance, over-stability of the contact discontinuity surface (e.g., to Kelvin-Helmholtz modes) may lead to nonlinear 
oscillations of the contact surface and the reconfinement shock itself that can enhance local dissipation 
in the region between the contact and the reconfinement shock (e.g., via generation of steepening waves, or by 
locally changing the incidence angle of the fluid upstream of the reconfinement shock), giving rise to acceleration 
of electrons and positrons to nonthermal energies.  Or it could be that a fraction of the particles are Fermi 
accelerated in the sheared flow as described in Ostrowski (1998). 
In order to account for such processes we suppose that a fraction $\xi_e$ of the thermal energy behind the 
reconfinement shock is injected as a power 
law distribution of electrons\footnote{The shape of the electron spectrum depends on the details of the acceleration 
mechanism at work.  Since we are merely interested in the net cooling rate the use of a single power law is sufficient.}:
$dn^\prime_e/d\epsilon^\prime_e\propto\epsilon^{\prime-2}_e$.  For simplicity we consider only synchrotron 
cooling of the nonthermal population.  Fixing $\xi_B$ and $\xi_e$ we then compute the synchrotron emissivity of the shocked 
gas, and incorporate it into the flow equations of the shocked layer.  
The details are given in the appendix.    To simplify the analysis further the fractions  $\xi_e$ and $\xi_B$ are taken to be 
constants along the channel (independent of z).  
The free parameters of the model are the power $L_j$, opening angle $\theta_j$, Lorentz factor $\Gamma_{j0}=\Gamma_j(z=z_0)$,
the mass flux $\dot{M}_j$ and the product $\xi_e\xi_B^{3/4}$ of the injected flow, and $p_0$, $\eta$ of the external medium.

Examples are shown in figures 1- 3.  In all the examples shown the confining medium is static with a pressure
profile $p_{ext}=p_0 (z/z_0)^{-\eta}$, as discussed above, and the jet power is $L_j=10^{44}$ erg/s.  The 
integration starts at $z_0=1$ pc, where the opening 
angle of the injected flow is taken to be $\theta=10^\circ$.  Integration of the full set of
equations confirms the scaling derived analytically in eq. (\ref{zstar}).
The effect of cooling is examined in fig. 1. In this example the ejected 
flow is assumed to be cold, in the sense that its energy density is dominated by rest mass energy. 
To be more precise, we take $L_j=\dot{M}_j\Gamma_{j0} c^2$ at the initial impact point $z=z_0$, 
(as stated above, in all cases considered here $\xi_B<<1$ and has a little effect on the
shock jump conditions that we ignore).   We considered both, baryonic jets and pure electron-positron jets and, 
as expected, found little dependence on the outflow composition, with the exception of the temperature behind
the shock which is higher in the baryonic case, owing to the lower density.  The difference between 
the three cases shown in fig. 1 is in the synchrotron cooling rate behind 
the oblique shock, which is controlled by the parameters $\xi_e$ and $\xi_B$.  The ratio of the total luminosity 
radiated away by the shocked jet layer,
\begin{equation}
L_c =\int_{z_0}^{z^\star}{\pi(r_c^2-r_s^2)S^0dz},
\end{equation}
where $S^0$ is given in Eq. (\ref{s0lab}) in the appendix, and total jet power $L_j$ in the upper, middle and lower panels 
is $L_c/L_j=0$ (no radiative losses), 0.1, and 0.27, respectively.  
As seen, in all cases the shock (indicated by the dashed line) approaches the axis
at the same distance from the injection point, consistent with the scaling in eq. (\ref{zstar}).  
The main effect of the radiative cooling, as illustrated in fig. 1, 
is to increase the shock compression ratio, thereby reducing the width of the shocked layer and, as a result, the 
cross-sectional radius of the jet at the point $z^\star$.  As demonstrated in the bottom panel, substantial 
focusing can be accomplished even for a modest radiative efficiency. We have also made some runs where only cooling of 
the thermal plasma has been incorporate into the model ($\xi_e=0$) and found the same behavior essentially.
Significant focusing occurs, as expected, when $t_{cool}\simlt t_f$ in Eq. (\ref{th-cool}).  In reality, the 
level of focusing may be limited by other components that have been neglected in our model, e.g., magnetic pressure (see 
further discussion below), or the pressure of nonthermal baryons accelerated by the sheared flow.

The effect of cooling on the structure of the flow is less dramatic in hot jets, in which the dimensionless enthalpy per particle 
of the unshocked fluid is much larger than unity (see fig. 2).  The reason is that in that case the bulk energy 
of the shocked fluid is dominated by pressure rather than rest mass energy and, therefore, a larger fraction of the total 
jet energy must be radiated away in order to significantly compress the shocked layer.  

Figure 3 exhibits solutions obtained for different profiles of the external pressure.  In this example we fixed the 
parameters $\xi_B$, $\xi_e$, $L_j$, $\dot{M}_p$ and $p_0$ and varied $\eta$.  Fixing $\xi_B$ and $L_j$ means  essentially 
that the magnetic field in the shocked layer is the same for all cases.   As expected from eq. (\ref{zstar}), the
location of the reflection point is closer for smaller values of $\eta$.  This is a consequence of the larger momentum transfer
across the interface separating the ambient medium and the jet, that forces larger 
deflection angles of streamlines crossing the shock.  Since the transverse component of the incident energy flux 
dissipates behind the shock this also means larger dissipation.  This is the reason why the Lorentz factor of the 
shocked fluid is smaller for smaller values of $\eta$, as seen in fig. 3.
The larger dissipation also leads to a much faster cooling and, as a result,
a much better focusing of the jet.  The fraction of the jet power that has been radiated away in each case is
indicated.

The above analysis neglects the contribution of magnetic pressure in the shocked jet layer.  This pressure 
may support the cooling layer and suppress focusing in cases where the field is predominantly poloidal 
or turbulent, and strong enough.  On the other hand, a toroidal component gives rise to 
magnetic hoop stress that actually helps collimating the flow.  The presence of a helical component has been
inferred from radio polarization maps in some blazars (e.g., Gabuzda et al. 2004).  At any rate,
we have verified that in all cases studied above the magnetic pressure was not important, except near the
point $z^\star$ in the upper panel in Fig 3. 

In steady state, reflections of the converging shock at $z^\star$ (where the shock crosses the axis)
generally leads to formation of conical shocks.  However temporal fluctuations may lead to more 
complicated structures, in particular formation of ``internal'' shocks in the vicinity of the recollimation nozzle.  To model
such effects requires 2D, time dependent simulations.  There is evidence for such features in 
the full 2D simulations performed by Alloy et al. (2005).  A considerable fraction of the remaining jet power 
may dissipate via these internal shocks in a region much smaller than what would be expected in the case 
of a conical jet.  This may lead to large amplitude variations over timescales much shorter than the jet radius,
as discussed further below.

An important question is whether the narrow structure formed beneath the reflection point is stable.   
Our model cannot account for any temporal effects, particularly those associated with Kelvin-Helmholtz
instabilities at the interface separating the shocked jet layer and the ambient matter.
If the instability grows to a nonlinear state over the expansion time then
pinching of the fast jet and mixing of ambient material with the
jet fluid near the interface is anticipated.  This can modify the
transverse structure of the jet.  Numerical simulations exhibit some
evidence for such instabilities (e.g., Aloy, 2005; Hardee \& Hughes
2003), but the details should depend on the configuration of the
confining medium, and on the structure of the viscous boundary layer.  Detailed analysis of the stability
of the cooling jets considered here requires 3D simulations with an appropriate setup.  

\section{The case of HST-1 in M87}
The rapid variability of the resolved X-ray emission from the HST-1 complex sets a limit
on the cross-sectional radius of the jet at the location of HST-1 that depends on the fraction of jet power 
radiated as X-rays (and TeV emission, if indeed originating from the same location).
Estimates of the jet power in M87 yield $L_j\simgt10^{44}$  erg s$^{-1}$ (Bicknell \& Begelman 1996).
For the observed X-ray (Cheung, et al. 2007) and TeV (Aharonian, et al. 2006) luminosities, 
$L_{\rm TeV}\sim L_{\rm x}\simlt10^{41}$ erg s$^{-1}$, this
implies a rather small radiative efficiency, $\epsilon_r\equiv L_{\rm x}/L_j\sim10^{-3}$.  In order to account for the observed luminosity the size scale of the fluctuations producing the X-ray flare must satisfy $d\simgt2\epsilon_r^{1/2} a$, where $a$ denotes the cross-sectional radius of the jet at the dissipation region.  The variability time, on the other hand,
constrain the size of these fluctuations to be $d<\delta_D ct_{\rm var}$.  Combining the two constraints yields 
$a<0.5\epsilon_r^{-1/2}\delta_D ct_{\rm var}$.  
The apparent speed measured for the superluminal sub-knots in the HST-1 complex, $\beta_{app}\sim4$,
implies $\Gamma>4$ for the fluid passing the HST-1 knot and a viewing angle $\theta_n\sim30^\circ$ .
Adopting $\theta_n=30^\circ$, $\Gamma=4$, and $\epsilon_r=10^{-3}$, we estimate
$a<1$ pc for the reported X-ray variability, consistent with the associated HST source, and
$a<0.05$ pc for the TeV variability if indeed associate with HST-1 (Cheung et al. 2007).  For a conical jet this requires an opening 
angle $\theta_j=a/z_{\rm HST1}<10^{-2}$ and $\theta_j<5\times10^{-4}$, respectively, that seem somewhat unrealistic.

\begin{figure}[h]
\centering
\includegraphics[width=11cm]{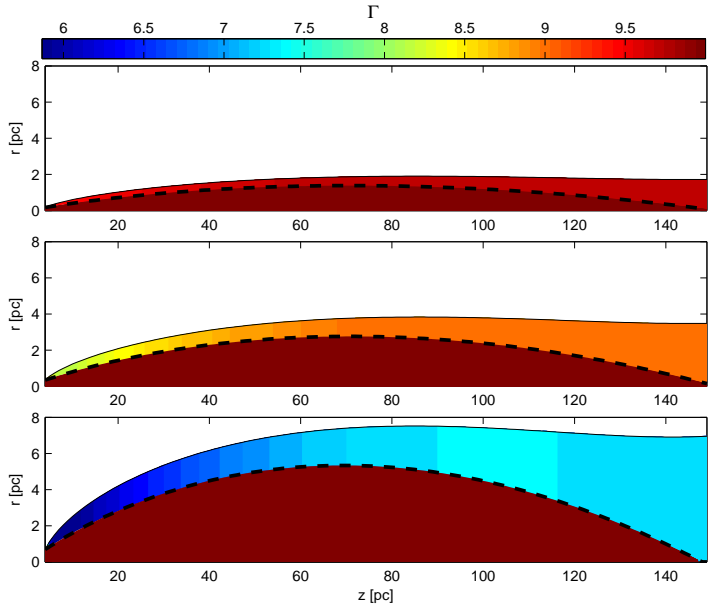}
\caption{\small{Lorentz factor of the shocked and unshocked layers of a baryonic jet in the case of confinement by a 
static condensation with a pressure profile $p_{ext}(z)=p_B(z/z_B)^{-\eta}$.   The integration starts at $z_0=4$ pc.
In all cases shown $L_j=5\times10^{43}$ erg s$^{-1}$, $\Gamma_{j0}=\Gamma_j(z=z_0)=10$, 
$z_B\sim200$ pc, $p_B=1.5\times10^{-9}$ dyn cm$^{-2}$ 
and $\eta=0.6$.  The semi-opening angle of the injected outflow is: $\theta_j=0.05$ (upper panel) 
$\theta_j=0.1$  (middle panel), and $\theta_j=0.2$  (bottom panel).}}
\end{figure}

ST06 proposed that HST-1 reflects the location of the point where the reconfinement shock reaches the axis.
The picture they envisaged is that, at a distance $z_0$ of several parsecs from the central engine the jet encounters
a gaseous condensation having a pressure profile of the form $p_{ext}(z)=p_B(z/z_B)^{-\eta}$, with 
the values $z_B\sim200$ pc, $p_B=1.5\times10^{-9}$ dyn cm$^{-2}$ and $\eta=0.6$ adopted in the reconfinement zone.
By employing a model developed by Komissarov and Falle (1997), they computed the profile of the converging 
reconfinement shock and found that it reaches the axis at a distance
\begin{equation}
z^\star=\left[\frac{(2-\eta)^2\xi_1}{4}\frac{L_j}{\pi c p_0z_B^\eta}\right]^{1/(2-\eta)}
\label{nozz}
\end{equation}
from the central engine, which for $\eta<2$ coincides with eq. (\ref{zstar}) in the limit $z_0\rightarrow0$.
For the parameters adopted by ST06 a total jet power of $L_j\simeq10^{44}$ ergs s$^{-1}$ is 
required to match $z^\star$ with the location of HST-1, consistent with the estimate of Bicknell \& Begelman (1996)
\footnote{ST06 imposed the additional restrictions that the kinetic pressure of the jet 
at $z=z_0$ equals the ambient pressure, viz., $p_j(z=z_0)=p_{ext}(z=z_0)$, and that the opening angle of the jet 
at $z=z_0$ equals the inverse of its Mach number.  This uniquely determines $L_j$ and $z^\star$.  Other situations can 
be envisaged that do not require such restrictions.}.
The treatment of ST06 is incomplete, as it does not compute the structure of the shocked jet layer, and in particular
the cross-sectional radius of the jet at the location of HST-1.  To study this 
we applied our semi-analytic model to M87.  Since only a small fraction of the total jet power is converted to radiation 
in this object, viz., $L_c/L_j<10^{-3}$, 
it is not expected to affect the dynamics of the system (see fig. 1).  We therefore set $\xi_e=0$.  We emphasis that even
though unlikely to affect the structure, synchrotron emission may still be observable from regions where the 
recollimation shock is sufficiently strong.

An example is shown in fig. 4, where the above values of the ambient gas parameters have been invoked.
In this example the integration starts at $z_0=4$ pc, where $\theta_j$ is fixed.  Starting the integration
at smaller distances $z_0$ does not change the solution much.  This is expected since the ratio of 
the outflow ram pressure and the external pressure increases as $z$ decreases (roughly as $z^{-1.4}$),
so that at $z<<z^\star$ the shocked jet layer is very thin and the jet profile is essentially unaffected by the ambient matter. 
As seen the location of the reflection point $z^\star$ is the same in all cases but the profile of the contact discontinuity depends on 
the opening angle of the jet at the initial location $z_0$.  It is also seen that collimation occurs on intermediate 
scales (20 to 40 pc for the parameters used in fig. 4) where the recollimation shock is strongest.  Observable 
synchrotron emission may be produced behind the shock in this region.  Beyond the collimation radius
the shock becomes much weaker and we expect little or no radiation there.  This appears to be consistent 
with morphology of the radio jet on sub-kpc scales.

\section{Discussion}
We have constructed a class of semi-analytical models for the collimation of a relativistic 
cooling jet by the pressure of a surrounding medium.  To illustrate the effect of cooling 
we assumed, for computational convenience, that radiative losses behind the collimation shock are dominated by synchrotron
emission of accelerated electrons.  However, the structure of the shocked jet layer depends
merely on the net cooling rate and not on the details of the cooling mechanism, provided
momentum losses are small.  The main conclusion is that, under certain 
conditions, radiative cooling of the shocked jet layer can lead to a good focusing 
of the jet.  Focusing requires ambient medium with sufficiently flat pressure profile, and is most effective 
when the jet is cold, that is, when its energy flux is carried predominantly by rest mass.  The reason is that in the latter
case modest radiative efficiency is already sufficient to reduce the temperature behind the 
collimation shock considerably, giving rise to a high compression of the shocked jet material.  
The location where the reconfinement shock reaches the axis depends solely on the external pressure
and the jet power.  However, the profile of the jet and, in particular, its cross-sectional radius at 
the reflection point depend also on its opening angle and specific enthalpy at the base of the 
reconfinement zone.  Our analysis does not account for the effect of radiative drag, that
might be important in cases where ERC emission dominates.  The latter may somewhat alter the 
dynamics and structure of the shocked layer, but should still lead to focusing under the same conditions.  In fact, 
we have shown that if the intensity of the background radiation field is as large as typically 
assumed in ERC models than cooling of the thermal electrons alone may be rapid enough to cause
substantial focusing.  Since the Lorentz factor of the flow near the contact discontinuity surface is smaller than 
that of the unshocked jet the emission from this region is expected to be less beamed. The GeV emission detected
recently by Fermi in the radio galaxies Cen A and NGC 1275 (Abdo, et al. 2009a,b) may originate from this region.
Temporal fluctuations of the central engine may lead to a more complex behavior.  In particular,
internal shocks may form below the point where the collimation shock intersect the axis and contribute to the emission.  There is some evidence for
such a behavior in BL Lac (Marscher et al. 2008).  

Radiative reconfinement can channel the outflow into a region having a very small cross-sectional radius.
Reflections of the converging recollimation shock there should give rise to additional dissipation 
via formation of internal shocks.  This can naturally explain ejections of 
superluminal radio sub-features and rapid variability from an otherwise quasi-stationary region, 
as occasionally observed in radio jets of blazars (Jorstad et al. 2001) and, most notably, in M87.
The pattern speed of the main radio knot may be associated with the location of the reflection point while the Lorentz factor 
inferred from the variability of the VHE emission is associated with the speed of the fluid passing that point.
The superluminal sub-features ejected from the main knot can be interpreted as reflections of the 
internal shocks that formed at the reflection point. 

The confining medium in blazars may be associated with matter in the broad line region.  The characteristic pressure
measured in broad line emitters is $p_{BLR}\sim10^{-2}-10^{-3}$ dyn cm$^{-2}$, on scales $z_{BLR}\sim 0.01 -1$ pc.
The nature of the broad line emitters is yet an open issue.  If consists of small clouds, 
as envisioned by some, than the question remains as to how these clouds are confined (if at all).  
Most likely, some intercloud medium with a similar pressure is present.  Choosing for illustration $p_0=p_{BLR}$, $z_0=z_{BLR}$ 
and $\eta=1$ in eq. (\ref{zstar}), we obtain: $z^\star\simeq\ 2.5
(L_j/10^{46} {\rm erg\ s^{-1}})(p_{BLR}/10^{-2} {\rm dyn\ cm^{-2}})^{-1}(z_{BLR}/0.1\ {\rm pc})^{-1}$ pc. 
The pressure distribution in the BLR is uncertain.  However, as illustrated in fig. 3 the location of the intersection
point, $z^\star$, does not depend strongly on $\eta$.  Thus, we conclude that reconfinement, and even substantial 
focusing if radiation losses are significant, can occur on VLBI scales, as, e.g., proposed for 3C454.3 by Sikora et al. (2008). 
For the above numbers we find that if about 30 percents of the bulk energy is radiated away behind the shock,
viz., $L_c/L_j\simeq0.3$, then the cross-sectional radius of the jet at $z^\star$ is $a\simeq10^{-2.5}z^\star$.

In M87 the radiative efficiency appears to be very low and cooling is unlikely to affect the structure of the 
confined jet.  However, the flat pressure profile of the ambient gas on sub-kpc scales (ST06) leads nonetheless
to extremely good collimation.  For the ambient pressure profile adopted by ST06 we find a cross-sectional radius
of $a/r_{HST}\simlt10^{-2}$ if the opening angle of the jet at the base of the reconfinement zone ($\sim$ several parsecs) satisfies
$\theta_j\simlt0.1$.  This level of collimation is sufficient to account for the X-ray variability observed, but not
for the variability of the TeV emission.  The latter would require extremely good collimation on sub-parsec scales by some other mechanism.
More likely, the TeV emission originated from a different location, e.g., the 
black hole magnetosphere (Levinson 2000; Neronov \& Aharonian 2007).

\section*{Appendix A: Inclusion of synchrotron cooling}
Let $T_{js}^{\mu\nu}$ denotes the stress-energy tensor of the shocked jet layer (including the nonthermal particles).
The dynamics of the shocked jet layer is governed by the continuity equation and by energy-momentum conservation: 
\begin{equation} 
\partial_\nu T_{js}^{\mu\nu}=S^{\nu},
\end{equation}
where $S^\mu$ denotes the source term associated with energy and momentum losses of the shocked fluid.
Our treatment assumes that energy losses are dominated by 
synchrotron emission of electrons having a power law energy distribution,  $n^\prime(\epsilon_e^\prime)=K\epsilon_e^{\prime-2}$,
with a lower cutoff $\epsilon^\prime_{e,min}$ which we take to be $m_ec^2+kT_{js}$ for convenience, where 
$T_{js}$ is the temperature of the shocked layer, and an upper cutoff
\begin{equation}
\epsilon^{\prime 2}_{e,max}=m_ec^2+\frac{3}{2}\frac{m_e^3c^6}{e^3}\left(\frac{\eta_{acc}}{B^\prime}\right),
\end{equation}
which was obtained by equating the synchrotron cooling rate 
with the acceleration rate, $t_{acc}^{-1}=\eta_{acc} eB^\prime/m_ec$, where $\eta_{acc}$ represents the acceleration efficiency, and $B^\prime$ 
the magnetic field, as measured in the fluid rest frame.
The power per unit volume emitted in the fluid rest frame 
can be readily computed using the synchrotron emissivity $j^\prime_{syn}(\nu)$.
In terms of the total energy density of nonthermal electrons, 
$u^\prime_e=\int_{\epsilon^\prime_{e,min}}^{\epsilon^\prime_{e,max}}\epsilon_e^\prime n^\prime(\epsilon_e^\prime)d\epsilon_e^\prime$,
and the magnetic field energy density, $u^\prime_B=B^2/8\pi$, it can be expressed as:
\begin{equation}
S^{\prime0}=-\int_{\nu^\prime_{th}}^{\nu^\prime_{max}}{j^\prime_{syn}(\nu^\prime)d\nu^\prime}=-\frac{6}{(8\pi)^{3/4}}
\sqrt{\frac{\eta_{acc}}{2}}\Gamma\left(\frac{7}{3}\right)
\Gamma\left(\frac{2}{3}\right)\frac{e^{5/2}u^\prime_eu_B^{\prime3/4}}{m^2c^3\ln(\epsilon^\prime_{e,max}/\epsilon^\prime_{e,min})},
\label{s0}
\end{equation}
where $\Gamma(x)$ is a Gamma function with argument $x$.  
As a closure condition for the set of hydrodynamic equations we take $u_B^\prime$ to be a fixed fraction $\xi_B$ of the total
energy density behind the shock and $u^\prime_e$ to be a fraction $\xi_e$ of the thermal energy behind the shock 
(which for a strong shock is roughly the energy dissipated in the shock). Specifically,
$u_B=\xi_Bw_{js}$, $u_e=\xi_e(w_{js}-n_{js}mc^2)$, where $n_{js}$ and $w_{js}$ are the proper baryon density and enthalpy of 
the shocked gas (see below).  We also adopt $\eta_{acc}=1$ for convenience.
Since the emission is isotropic in the rest frame of the fluid there are no momentum losses
there, and so $S^{\prime \mu}=(S^{\prime0},0,0,0)$.  Transforming to the star frame we then obtain 
\begin{equation} 
S^\mu=\Lambda^{\mu}_{\nu}S^{\prime\nu}=\Gamma_{js} S^{\prime 0}(1,{\bf \beta}_{js})
\label{s0lab}
\end{equation}
Here $c\beta_{js}$ is the 3-velocity of the shocked fluid and $\Gamma_{js}$ the corresponding Lorentz factor. 
Using the above derivation, eqs (22) of BL07 is generalized to: 
\begin{equation}
\frac{d}{dz}\left[w_{js}\Gamma_{js}U_{js}(r_c^2-r_j^2)\right] - S^{\prime0}\Gamma_{js}(r_c^2-r_j^2)=
2w_j\Gamma_jU_jr_j\frac{\sin\delta_j}{\cos\alpha_j},\label{energy_eq_j}
\end{equation}
and eq. (24) to:
\begin{equation}
\frac{d}{dz}\left[w_{js}U_{js}^2(r_c^2-r_j^2)-r_j^2p_{js}\right]+r_c^2{dp_{js}\over dz}- S^{\prime0}U_{js}(r_c^2-r_j^2) = 
2w_jU_j^2r_j\frac{\cos\theta\sin\delta}{\cos\alpha} - p_j{dr^2_j\over dz},\label{momZ_eq_j}
\end{equation}
with $S^{\prime0}$ given by eq. (\ref{s0}).  The various quantities are defined in BL07.  The 
total energy density of the shocked layer is the sum of the the contributions of the thermal 
plasma, the nonthermal (accelerated) particles, rest mass and magnetic fields, viz., $u_{js}=n_{js}m_pc^2+u_{gas}+u_e+u_B$.
(In all cases studied above the magnetic field is dynamically unimportant, viz., $\xi_B<<1$, so that $u_B$ and the 
magnetic pressure $p_B$ can be practically neglected in the flow equations.)
Since the nonthermal particles are relativistic their pressure 
is given by $p_e=u_e/3$.  The pressure of the thermal plasma is related to its energy density through 
$p_{gas}=(\gamma-1)u_{gas}$, where $\gamma$ is the adiabatic index associated with the thermal gas that 
we take to be either $\gamma=4/3$ or $\gamma=5/3$, depending upon whether the temperature of the thermal plasma is 
relativistic or not.   The enthalpy of the shocked gas is given by,
\begin{equation}
w_{js}=n_{js}m_pc^2+\gamma/(\gamma-1)p_{gas}+4p_e.
\label{enthalp}
\end{equation} 
The total pressure of the shocked layer is the sum $p_{js}=p_e+p_{gas}$, and must satisfy
$p_{js}(z)=p_{ext}(z)$ by virtue of momentum balance across the contact discontinuity surface (BL07).
With the above parametrization for $u_e$ we obtain, using Eq. (\ref{enthalp}),
\begin{equation}
u_e(z)=\frac{3\gamma\xi_e}{\gamma\xi_e+(\gamma-1)(3-4\xi_e)}p_{ext}(z).
\label{pe}
\end{equation} 
Eqs. (\ref{s0}) - (\ref{pe}) augmented by the continuity equation (eq. 20 in BL07), the normal component of the momentum 
equation (eq. 26 in BL07), and the condition $p_{js}(z)=p_{ext}(z)$ form a closed set for the unknown variables 
of the shocked fluid.

We thank C. Dermer for comments, and the anonymous referee for constructive criticism. 
This work was supported by an ISF grant for the Israeli Center for High Energy Astrophysics.

\end{document}